\documentclass[jap,twocolumn, showpacs,preprintnumbers,amsmath,amssymb,superscriptaddress,floatfix]{revtex4-1} 
\usepackage[]{graphicx}
\usepackage{graphics}
\usepackage{subfigure}
\usepackage{amsmath}
\usepackage{tabularx}
\usepackage{bm}
\usepackage{multirow}

\begin{document}

\title[Magnetoplasmonic design rules for active magneto-optics]{Magnetoplasmonic design rules for active magneto-optics}

\author{Kristof Lodewijks}
\homepage[Website: ]{http://www.kristoflodewijks.be}
\email[E-mail: ] {lode@kristoflodewijks.be}
\affiliation{Department of Applied Physics, Chalmers University of Technology, 41296 Gothenburg, Sweden}
\altaffiliation{K. L. \& N. M. Contributed equally to this work}

\author{Nicol\`{o} Maccaferri}
\email[E-mail: ] {n.maccaferri@nanogune.eu}
\affiliation{Nanomagnetism Group, CIC nanoGUNE Consolider, 20018 Donostia-San Sebastian, Spain}
\altaffiliation{K. L. \& N. M. Contributed equally to this work}

\author{Tavakol Pakizeh}
\affiliation{Faculty of Electrical Engineering, K.N. Toosi University of Technology, Tehran 16314, Iran}

\author{Randy K. Dumas}
\affiliation{Department of Physics, University of Gothenburg, 41296 Gothenburg, Sweden}

\author{Irina Zubritskaya}
\affiliation{Department of Applied Physics, Chalmers University of Technology, 41296 Gothenburg, Sweden}

\author{Johan \AA kerman}
\affiliation{Department of Physics, University of Gothenburg, 41296 Gothenburg, Sweden}
\affiliation{Materials Physics, School of Information and Communication Technology, KTH Royal Institute of Technology, 16440 Kista, Sweden}

\author{Paolo Vavassori}
\affiliation{Nanomagnetism Group, CIC nanoGUNE Consolider, 20018 Donostia-San Sebastian, Spain}
\affiliation{IKERBASQUE, Basque Foundation for Science, 48011, Bilbao, Spain}

\author{Alexandre Dmitriev}
\email[E-mail: ]{alexd@chalmers.se}
\affiliation{Department of Applied Physics, Chalmers University of Technology, 41296 Gothenburg, Sweden}

\begin{abstract}
Light polarization rotators and non-reciprocal optical isolators are essential building blocks in photonics technology. These macroscopic passive devices are commonly based on magneto-optical Faraday and Kerr polarization rotation. Magnetoplasmonics -- the combination of magnetism and plasmonics -- is a promising route to bring these devices to the nanoscale. We introduce design rules for highly tunable active magnetoplasmonic elements in which we can tailor the amplitude and sign of the Kerr response over a broad spectral range. 
\end{abstract}

\maketitle

Photonics uses the polarization of light as an information carrier in optical communications, sensing and imaging. Optical components that are able to manipulate the polarization such as rotators and non-reciprocal isolators are ubiquitously\cite{NatCommShalaby2013,NatPhotTamagnone2014,NatPhotYu2009,NatPhotonBi2011,StadlerIEEE2014}. Similarly, the state of polarization plays an important role in the photonic transfer of quantum information \cite{NatPhotonNorthup2014,NaturePeyronel2012}. The difficulty for the integration of the polarization-modulation components into waveguide optics or, even more prohibitive, into nanophotonic circuitry, is their macroscopic dimensions. Such large components are needed for polarization modification, for example, via magneto-optical Kerr and Faraday polarization rotation with magnetic bias. Thus, there are significant and growing efforts to implement new conceptual designs for efficient polarization control of propagating optical modes with planar plasmonic metasurfaces  \cite{NatMaterYu2014,ScienceKildishev2013,NLyu2012,PRLmonticone2013,IEEEholloway2012,ScienceAieta2013,SciencePendry2006}, or new materials like graphene to achieve future ultra-flat devices \cite{PRLgullans2013, NatNanoJu2011,PRBfallahi2012,ACSnanoFang2013,NatPhotTamagnone2014}. 

Within the paradigm of traditional magneto-optic materials such as metallic ferromagnets, the rapidly developing field of magnetoplasmonics merges the concepts from plasmonics and magnetism to realize novel and unexpected phenomena and functionalities for the manipulation of light at the nanoscale \cite{AOMarmelles2013,JOAarmelles2009,NatCommChin2013,NatPhotonTemnov2012,NatPhotonTemnov2010,BelotelovNatNano2011,BelotelovNatComm2013,TemnovNatComm2013,Gonzalez-DiazSmall2008,BanthiAdvMat2012,CtistisNL2009,PapaioannouOE2011,RazdolskiPRB2013,ValevACSnano2010}. Plasmon resonances, which are light-induced collective electron oscillations, enable energy confinement at the nanoscale, greatly enhancing the electromagnetic near-field at the resonant wavelength. Due to such strong field localization, plasmon resonances also show enhanced interactions with external magnetic fields, resulting in enhanced values of the magneto-optical Kerr signals in magnetoplasmonic materials. This allows exploring the non-reciprocal propagation of light by designing custom nanoscale magnetoplasmonic elements \cite{NatCommChin2013}. The majority of studies in magnetoplasmonics commonly combine (ferro-) magnetic materials with noble metals \cite{JOAarmelles2009,JOAarmelles2009,Gonzalez-DiazSmall2008,BanthiAdvMat2012} in order to explore strong magnetic behavior together with reduced damping of plasmonic modes in noble metals. Noble metals such as gold in themselves display sizeable magnetoplasmonic effects \cite{PineiderNL2013,PRLsepulveda2010,ACSnanoTuboltsev2013}, although high magnetic fields are required. It was recently shown that simple ferromagnetic nanoparticles support strong localized surface plasmon resonances (LSPRs) \cite{SmallChen2011} with reasonably small damping. Moreover, the plasmon resonances in these nanostructures can tune the Kerr rotation and ellipticity and even induce its controlled sign inversion \cite{NLbonanni2011,PRLmaccaferi2013}.

\begin{figure*}[!ht]
\includegraphics[width=12cm]{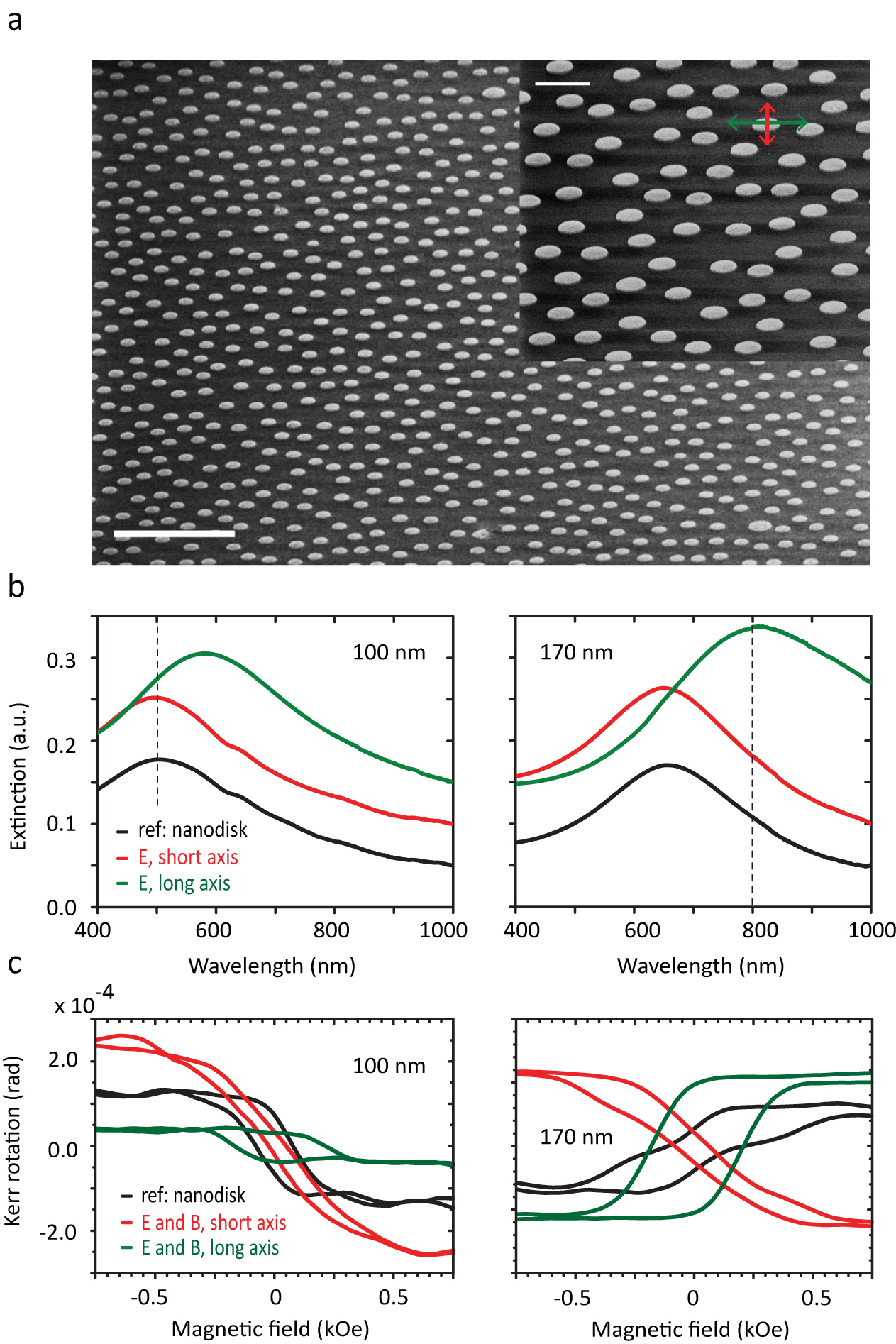}
\caption{(a) 60-degree tilted view scanning electron microscope (SEM) image of 170/240nm (short / long axis) short-range ordered nano-elliptical antennas, scale bar $2 \mu m$. Inset: closer view, scale bar $500 nm$. Incoming light polarization in (b) marked here with green and red arrows. (b) Optical extinction for different nanoantennas sizes (100 nm or 170 nm short axis on left and right panels, respectively) with the incident beam polarized along the different symmetry axes (green and red). (c) Measured L-MOKE spectra in p-polarization for 100/170 nm (left/right panel) nanostructures at 500 nm and 800 nm (dashed lines in (b)), respectively.}
\label{figure1}
\end{figure*}

Here we introduce the rules necessary to engineer nanoscopically thin and active metasurfaces of magnetoplasmonic nanoantennas. By active we refer to the ability of such a magnetoplasmonic metasurface both to multifold enhance magneto-optical polarization rotation over a broad spectral range and also the ability to change its sign by engineering the nanoantenna design. We first employ an archetypical plasmonic and magnetic nanostructure with geometrical anisotropy €-- a planar nanoellipse -- to devise a simple general concept for the broadband control of phase and amplitude, i.e polarization and/or intensity of light. The key to this functionality lies in the interplay of the directly excited and the magneto-optically induced dipoles in the individual magnetoplasmonic nanoantenna. We later extend this approach to fully three-dimensional (3D) ferromagnetic nanoantennas, where the combination of resonant and off-resonant nanoplasmonic modes further allows optical loss control in the magnetoplasmonic metasurfaces, while at the same time maintaining the active operation modality -€" something crucial when considering practical designs for polarization rotators. To stress the generality, we conceptualize the introduction of active magneto-optics in all common magneto-optical Kerr effect (MOKE) geometries -€" namely longitudinal (L-MOKE), polar (P-MOKE) and transverse (T-MOKE) \cite{IEEEfreiser1968}.

We tackle the emergence of active magneto-optics using a metasurface of nickel nanoantennas with various degrees of geometrical anisotropy, starting from circular nanodisks followed by nanoellipses with a fixed aspect ratio of 1.4 between the long and short axis. Our experimental system is a short-range ordered arrangement of nanostructures fabricated on a glass substrate ($\sim cm^2$) by hole-mask colloidal lithography \cite{AdvMatFredriksson2007} (Figure 1a). The nanoantennas are fabricated using polystyrene beads of 100 and 170 nm nominal diameters, resulting in circular nanostructures with 100 and 170 nm diameters and elliptical nanostructures with 100/140 nm and 170/240 nm ratios for the short/long axis respectively. The latter two are referred to as ``100 nm'' and ``170 nm'' nanoellipses in the discussion that follows. The thickness of all nanoantennas and reference films is 30 nm. 

From a purely optical standpoint it is wellknown that the polarizability is different along the various symmetry axes of anisotropic plasmonic nanostructures \cite{APLwang2012}. For all nickel nanoantennas we observe broad electric dipole resonances in the visible and near-infrared spectral range (Figure 1b). As expected, the spectral position for the dipole resonance in nanodisks and nanoellipses along the short axis are fairly similar, although the latter shows a minor blue shift and increased extinction. The resonance along the long axis is observed at larger wavelengths and shows further increased extinction. It is worth to point out, that even though the samples exhibit short-range order, the inter-particle spacings are sufficiently large to consider them as non-interacting. 

Magnetically, in contrast to nanodisks, nanoellipses develop in-plane easy and hard magnetization axes due to the magnetostatically induced shape anisotropy \cite{PRBvavassori2004}, which is observed in the L-MOKE magnetization loops for p-polarized light incident at $25^o$ in figure 1c (see also supporting information). Here the coercive fields and saturation fields for the nanodisks (black lines) and nanoellipses (red and green lines) directly reflect the effects of magnetic shape anisotropy. It is noteworthy to point out already here that there is a large tunability of the sign and amplitude of the Kerr rotation. For the case of 100 nm nanodisks and nanoellipses, excited at 500 nm (dotted vertical line in Fig. 1b, left panel), we observe the same sign for all traces, while the amplitudes markedly differ (Fig. 1c, left panel). In the case of 170 nm nanostructures, excited at 800 nm (dotted vertical line at Fig. 1b, right panel), we reach the opposite sign but equal amplitude of the polarization rotation, depending on the symmetry axis along which we probe the nanoantennas (Fig. 1c, right panel). 

To systematically demonstrate polarization rotation control in 2D nanoantennas, we plot the spectrally resolved polarization rotation for p-polarized light incident at $25^o$ at magnetic saturation (magnetic field above 1000 Oe) for different sample orientations and compare it with the intrinsic response of a continuous nickel film of the same thickness (Fig. 2). A very broadband spectral tunability emerges, both in sign and amplitude (see the raster bars in Figure 2a and 2b), which is most pronounced for the 170 nm nanoellipses at 800 nm. We observe that the polarization rotation displays the same amplitude but opposite sign when we excite it along the long or short axis of the elliptical nanoantenna. Conversely the reference nickel film shows zero polarization rotation at this wavelength (Fig. 2b). The inversion (zero) points for the polarization rotation are shifted with respect to the reference film in such a way that $\lambda_{inv,long} < \lambda_{inv,round} < \lambda_{inv,short}$. For the magnetoplasmonic nanoantennas these inversion points are observed close to the spectral positions of the in-plane plasmonic dipole perpendicular to the excitation direction (along y-axis). The insets show how the polarization rotation inversion can be continuously engineered to values between both extremes by simply rotating the nanoellipse antennas (in this example at $45^o$, orange curve). Comparison of the measured spectra with theoretical calculations show an outstanding agreement (see Figures 2c and 2d). The calculated spectra are obtained by computing the polarizability of an individual nickel nanoparticle (material parameters from reference \cite{JMMMvisnovsky1993}), taking into account the dynamic depolarization fields, as described in detail in references\cite{OptExprMaccaferri2013,PRLmaccaferi2013}. The nanostructures layer is modelled using a Maxwell-Garnett effective medium approximation (EMA) where the polarizability of the layer is calculated based on the response of the individual nanoparticles and taking into account their density. The effect of the substrate \cite{PSSmaccaferi2014} in the far-field MOKE-response is included using the transfer matrix method (TMM).   

\begin{figure*}[!ht]
\includegraphics[width=12cm]{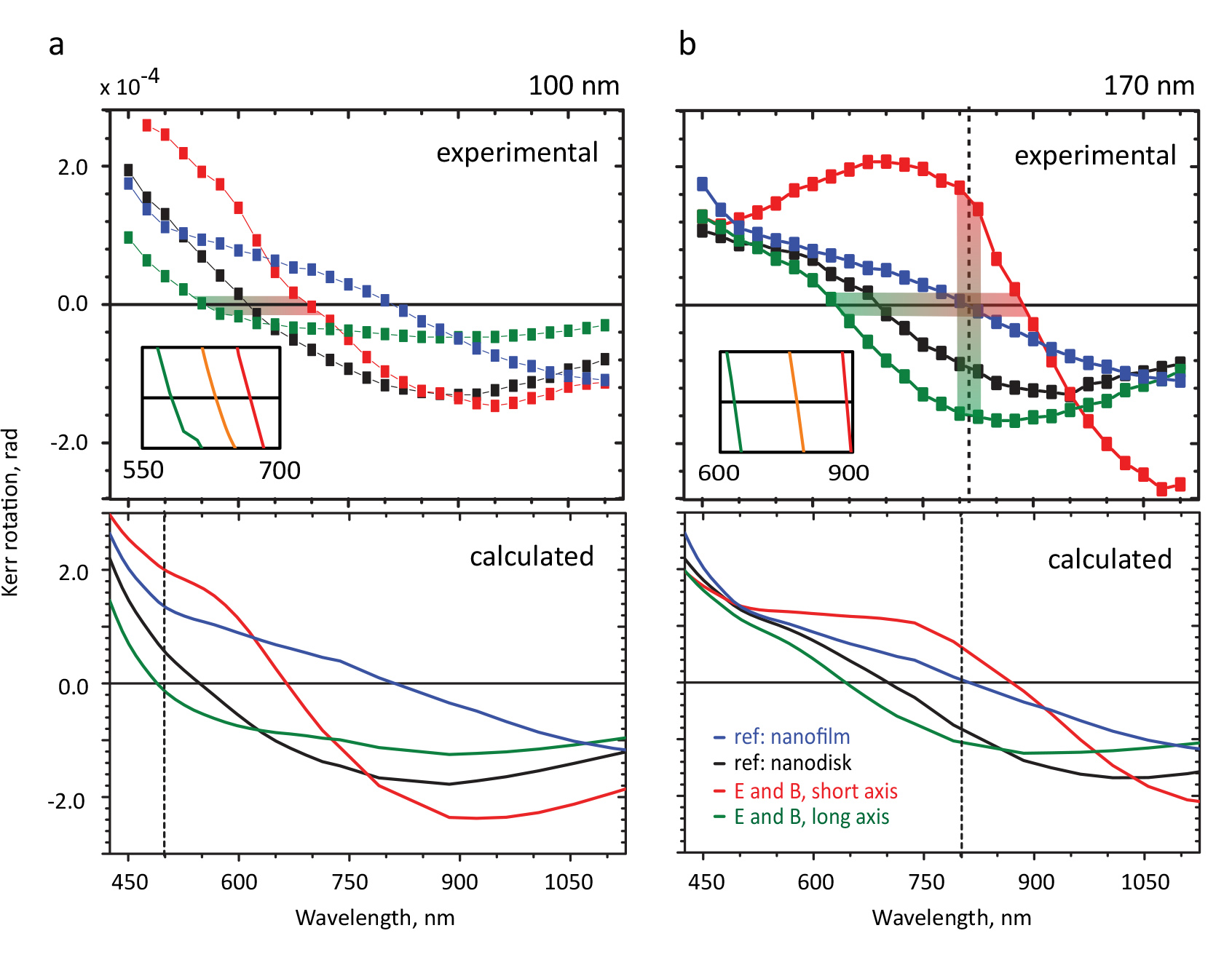}
\caption{Spectral dependence of the Kerr polarization rotation for the nanoantennas in p-polarization. Experimental MOKE spectra for 100 nm (a, top) and 170 nm (b, top) nanodisks and nanoellipses, along with the corresponding calculated ones, (a, bottom) and (b, bottom), respectively. The insets in (a, top) and (b, top) show the possibility for continuous control of Kerr rotation zeroing along both symmetry axes and for the intermediate case of $45^o$ in-plane sample rotation (orange). Raster bars in (a, top) and (b, top) show the tunability range of the Kerr polarization rotation in sign and amplitude.}
\label{figure2}
\end{figure*}

The broadband control of sign and amplitude of the Kerr polarization rotation originates from the off-diagonal terms of the dielectric tensor of nickel (intrinsic magneto-optical response)

\begin{equation}
 \tilde{\epsilon} = 
 	\begin{pmatrix} 
		\epsilon & \epsilon_{xy} & \epsilon_{xz} \\ 
		\epsilon_{yx} & \epsilon  & \epsilon_{yz} \\ 
		\epsilon_{zx} & \epsilon_{zy} & \epsilon \\ 
	\end{pmatrix} =
	\begin{pmatrix} 
		\epsilon & -i Q \epsilon m_z & i Q \epsilon m_y \\ 
		i Q \epsilon m_z & \epsilon  & -i Q \epsilon m_x \\ 
		- i Q \epsilon m_y & i Q \epsilon m_x & \epsilon \\ 
	\end{pmatrix}
\end{equation}

and the spectral dependence of the polarizability of the magnetoplasmonic nanoantennas. For the different MOKE geometries, different off-diagonal terms are activated by magnetizing the particles along different symmetry directions, namely along x (L-MOKE), y (T-MOKE) and z (P-MOKE), which corresponds to off-diagonal terms in yz, xz and xy respectively. Large differences in the polarization rotation amplitudes and signs for slight variations of nanoantenna designs are produced, indicating that controlling the nanoantenna shape opens up novel schemes and possibilities to finely tailor the polarization rotation at will. In the particular case of 2D nanoantennas in L-MOKE, the tunability originates from the interplay of the two in-plane plasmonic dipoles ($p$), which can be described by the nanoparticle polarizabilities ($\alpha_{ii}$, with $i=x,y,z$) along the different symmetry axes.  As we illuminate the particles with p-polarized light, only the in-plane mode along the x-direction ($p_x = \alpha_{xx} E_x$) is resonantly excited by the incident electromagnetic wave, as shown in the schematics in Figures 3a. At the same time there is a z-axis electric field component (due to the $25^o$ angle of incidence) which polarizes the nanoparticle non-resonantly ($p_z = \alpha_{zz} E_z$). Although the out-of-plane resonance of the nanoparticles occurs in the UV spectral region and thus does not contribute to the observed extinction / absorption from the nanoantennas, the weak polarizability in the visible is sufficient to couple efficiently through the spin-orbit interaction to the second LSPR mode in the nanoellipse plane along the y-direction such that \cite{OptExprMaccaferri2013}:

\begin{equation}
p_y = \alpha_{yz} E_z = \frac{\epsilon_{yz} \alpha_{yy} \alpha_{zz}}{(\epsilon - \epsilon_m)^2} E_z = \alpha_{so} \alpha_{yy} \alpha_{zz} E_z
\end{equation}

where $\epsilon$ and $\epsilon_{yz}$ are the diagonal and off-diagonal terms of the dielectric tensor of nickel $\tilde{\epsilon}$, respectively, $\epsilon_m$ is the dielectric function of the surrounding medium and $\alpha_{so} = \frac{\epsilon_{yz}}{(\epsilon - \epsilon_m)^2}$ is the spin-orbit induced off-diagonal term in the dielectric tensor, which is an intrinsic material property of nickel. We refer to this mode as the spin-orbit induced plasmonic dipole in the remainder of the discussion. As these two in-plane modes are spectrally well separated, and therefore give rise to a different response both in phase and amplitude of the polarizability, they offer us the possibility to obtain a very broadband tunability of the polarization rotation by 2D nanoantennas, which is further substantially altered compared to the reference film. Indeed, the polarization state of the reflected light here can simply be extracted from the ratio of the spin-orbit induced dipole $p_y$ and the two directly excited plasmonic dipoles $p_x$ and $p_z$

\begin{equation}
\begin{split}
 \frac{p_y}{p_x+p_z} 	& \propto \frac{p_y}{p_x cos{\gamma} +p_z sin{\gamma}} \\
					& \propto \frac{\alpha_{so} \alpha_{yy} \alpha_{zz} E_z}{cos{\gamma} \alpha_{xx} E_x + sin{\gamma} \alpha_{zz} E_z}\\ 
					& \propto \frac{\alpha_{so} \alpha_{yy} \alpha_{zz} \sin{\gamma} E}{\alpha_{xx} \cos^2{\gamma} E + \alpha_{zz} \sin^2{\gamma} E} \\
					& \propto \frac{\alpha_{so} \alpha_{yy} \alpha_{zz} \sin{\gamma}}{\alpha_{xx} \cos^2{\gamma} + \alpha_{zz} \sin^2{\gamma}} \\
\end{split}
\end{equation}

which contains the polarizabilities associated with all three modes of the nanoantenna. The incident angle $\gamma$ defines the coupling efficiency to the different modes and the radiation direction into the far-field. In terms of amplitude, we can see that both in-plane resonances occur in our spectral measurement window, while the out-of-plane plasmon mode occurs in the UV spectral region due to the limited nanoantenna thickness, resulting in extremely small amplitudes for $\alpha_{zz}$. Therefore, we can approximate the term in $\alpha_{zz} \approx 0$ in the denominator of equation 3 and rewrite the Kerr signals for 2D nanoantennas as

\begin{equation}
\frac{p_y}{p_x} \approx \frac{\alpha_{so} \alpha_{yy} \alpha_{zz} \sin{\gamma}}{\alpha_{xx} \cos^2{\gamma}}
\end{equation}

such that the amplitudes of the polarization rotation scale with the amplitudes of the polarizabilities of the nanoantenna modes along y and z and inversely with the amplitude along x. In terms of phase we can approximate the different contributions as well from this expression, where $E_x$ and $E_z$ are in phase

\begin{equation}
\begin{split}
\phi(\frac{p_y}{p_x+p_z}) 	& \approx \phi(\frac{p_y}{p_x})    \\	
						& = [\phi(\alpha_{yy})-\phi(\alpha_{xx})] +\phi(\alpha_{so}) + \phi(\alpha_{zz}) \\
						& = [\phi(\alpha_{yy}) + \phi(\alpha_{so})] - \phi(\alpha_{xx}) + \phi(\alpha_{zz})\\
\end{split}
\end{equation}

and in which the contribution of $\phi(\alpha_{zz})$ can be ignored as the out-of-plane resonance falls far outside of the measurement window. Deeper insight into the active polarization rotation rules emerges when we take a closer look at the phase of the nanoantenna mode polarizabilities. Schematics within Figures 3a demonstrate how the incident wave couples to two direct plasmon dipolar modes $p_x$ (resonant) and $p_z$ (non-resonant) for excitation along the short and long axes, respectively. For 2D nanoantenna, the non-resonant dipole mode  $p_z$ is coupled by the spin-orbit interaction to the spin-orbit induced plasmon dipole $p_y$ (resonant), which is now along the long or short axes in the respective cases. 

\begin{figure*}[!ht]
\includegraphics[width=12cm]{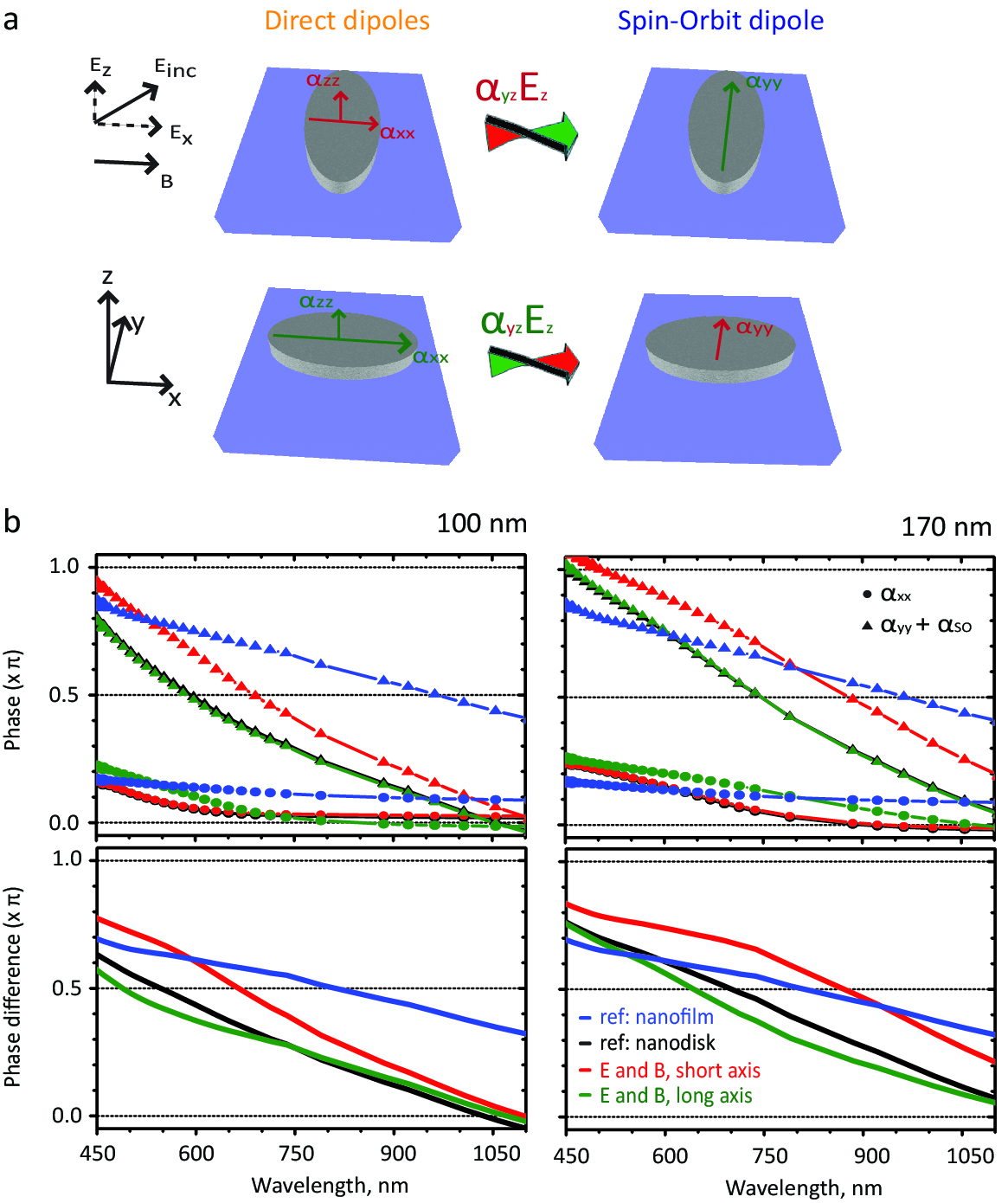}
\caption{(a) Schematic overview of the interplay of the directly excited (left, resonant $\alpha_{xx}$ and off-resonant $\alpha_{zz}$) and spin-orbit coupled (right, resonant $\alpha_{yy}$) nanoantenna modes that all together deliver exceptional light polarization rotation control. Light incidence (p-polarized) and applied magnetic field are marked. (b) Phase contributions (top row) of different polarizability terms for 100 nm (left) and 170 nm (right) nanoelliptical antennas along with the resulting phase differences in Fresnel coefficients (bottom row) for nanoantennas in (a).}
\label{figure3}
\end{figure*}

From the second line in Equation 5 we can see how the phase difference between both of the surface-plane plasmon dipole modes $[\phi(\alpha_{yy})-\phi(\alpha_{xx})]$ defines the broadband tunability on top of the intrinsic spin-orbit-induced ferromagnetic response of the nanoantenna, shifting the polarization rotation up (short axis) or down (long axis) compared to the reference ferromagnetic film. The different phase contributions can then be rearranged as in the third line of Equation 5, where $[\phi(\alpha_{yy}) + \phi(\alpha_{so})]$ indicates the sum of spin-orbit intrinsic material response $\alpha_{so}$ and the spin-orbit-induced magnetoplasmonic dipole mode $\alpha_{yy}$, perpendicular to the excitation direction. In Figure 3b we plot these calculated phase components for the far-field response, as matched with the experiment in Figure 2. The plotted phases are those of the Fresnel reflection coefficients $r_{pp}$ (circles) and $r_{sp}$ (triangles), with their ratio measured from the L-MOKE signal. The former is defined by the direct dipole with polarizability $\alpha_{xx}$ while the latter combines the contributions of the intrinsic material parameters $\alpha_{so}$ and the spin-orbit-induced dipolar mode $\alpha_{yy}$. Similar trends are observed for both sizes of nanoellipses, but become more pronounced with the increased degree of nanoantenna anisotropy -- compare left and right panels in Figure 3b. In terms of the direct dipole mode with $\alpha_{xx}$, we observe almost identical phases for nanodisks (black circles) and nanoellipses, excited along the short axis (red circles), while for excitation along the long axis (green circles) we see a similar trend at red-shifted wavelengths. However, for the spin-orbit-induced part, we observe a significantly different trend, as in this case the spectral response is dominated by the induced dipole mode with $\alpha_{yy}$, which determines the shift compared to the intrinsic material response $\alpha_{so}$ (film case, blue triangles). We see that the spin-orbit-induced phase terms are nearly identical for the nanodisks (black triangles) and the long-axis-excited nanoellipses (green triangles), as in this particular configuration the spin-orbit induced plasmon mode is along the short axis (see the model in Fig. 3a, bottom). For the short-axis excitation case (red triangles) we now observe a similar trend at red-shifted wavelengths, as the spin-orbit-induced LSPR is now along the long axis (model in Fig. 3a, top). By taking the difference between the spin-orbit induced dipole terms (solid lines with triangles in Fig. 3b, top) and the directly excited dipole terms (solid lines with circles in Fig. 3b, top), we obtain the phase contributions to the spectral Kerr polarization rotation, which are shown in Fig. 3b, bottom. We see that when the phase difference between the two components equals $\pi / 2$ the polarization rotation vanishes (see Fig. 2b, when Kerr polarization rotation crosses zero and compare with Fig. 3b, bottom), and that the relative difference compared to the intrinsic spin-orbit contribution determines the direction of the polarization rotation shift for a given nanoantenna with respect to the thin film. The absolute value of the polarization rotation then relates directly to the amplitude of the polarizability of the two surface-plane localized dipole modes, and the relative phase difference between them. 

In the previous section we illustrated how broadband control of the Kerr polarization rotation is generated in 2D nanoantennas, probed with p-polarized L-MOKE. In earlier work on similar nanostructures in P-MOKE \cite{PRLmaccaferi2013}, we observed polarization rotation values about one order of magnitude larger, but with substantially less tunability, due to the fact that the MOKE response only relies on one spin-orbit induced plasmon mode, as opposed to a direct and a spin-orbit induced dipole mode in L-MOKE. Here we generalize the concept of active magneto-optics by magnetoplasmonic design to all three conventionally employed MOKE geometries, namely L-MOKE, P-MOKE and T-MOKE, by designing a set of specially-configured 3D nanoelliptical antennas of similar dimensions (Figure 4a). We introduce the nanoantennas this time with {\it two resonant} plasmon modes in the visible spectral range (i.e., with elements sized 170 and 240 nm) and one off-resonant mode (element sized 30 nm). See also in Figure 4a (lower row) the sets of corresponding nanoantennas supporting dipolar plasmon modes (resonant - orange/yellow, off-resonant - grey). First we check the L-MOKE for all nanoantenna configurations in Figure 4b (top) - data colors corresponding to the color markers / nanoantennas configurarions in Figure 4a, where we observe much larger polarization rotations (compare with the absolute polarization rotations in Figure 2) for the cases where $\alpha_{xx}$ is off-resonant (red and black) and the spin-orbit coupled modes $\alpha_{yy}$ and $\alpha_{zz}$ are resonant. For the other two configurations (green and blue), on the other hand, we observe much smaller Kerr rotation values, which are similar in amplitude to the 2D case. This can be understood from the 3D nanoantennas geometry, as now one of the spin-orbit coupled modes $\alpha_{yy}$ is off-resonant, similar to the 2D case where $\alpha_{zz}$ was off-resonant. From equation 3, it is immediately clear that in the former case the numerator is large and the denominator - relatively small, while in the latter case the numerator is relatively small and the denominator is large, explaining the observed trends. It is worth noting that for 3D nanoantennas we can hardly approximate the phase with a simplified expression as in Equation 5. 

\begin{figure*}[!ht]
\includegraphics[width=10.5cm]{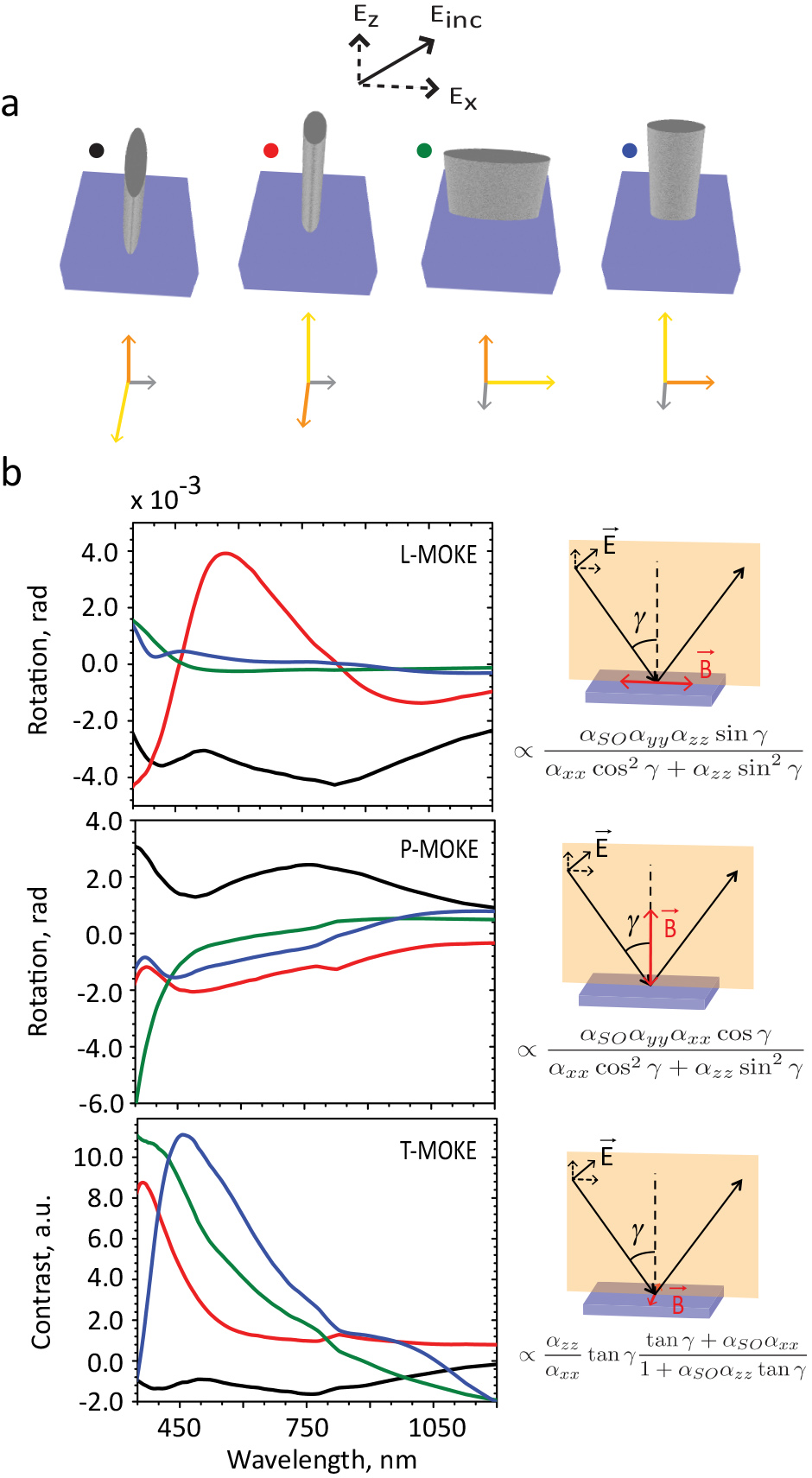}
\caption{(a) Schematic set of 3D nanoantennas (top row) with highlighted dipolar on (yellow, orange) / off (grey) resonance modes (bottom row). p-polarized light is always used. (b) Calculated Kerr signals for the nanoantennas in (a) in L-MOKE (top, rotation), P-MOKE (middle, rotation) and T-MOKE (bottom, contrast). Right panels feature schematics of the corresponding MOKE configurations, along with the analytical expressions for the observed polarization rotations.}
\label{figure4}
\end{figure*}

Figure 4b (middle) features the calculated P-MOKE polarization rotations and Figure 4b (bottom) - T-MOKE signals. For both we observe substantial differences between the various nanoantenna configurations, which can again be understood from the analytical expressions for the polarization rotations, featured next to the schematics of MOKE geometries in Figure 4b (detailed derivations are available in Supporting information). In P-MOKE (Figure 4b, middle) we observe the largest polarization rotation when $\alpha_{xx}$ is off-resonant and $\alpha_{yy}$ and $\alpha_{zz}$ are resonant. Due to the fact that only one of the polarizabilities in the numerator is resonant while the denominator is the same as for the L-MOKE case, here in P-MOKE the absolute rotation values are smaller than for L-MOKE (compare red and black curves for the two cases). The $\cos{\gamma}$ term in the numerator compensates for the effect of the weaker polarizability in $\alpha_{xx}$ (as compared to $\sin{\gamma}$ for L-MOKE). This is a striking result, as typical passive magneto-optical thin films produce overwhelmingly larger polarization rotation in P-MOKE as compared to L-MOKE. This illustrates the tremendous opportunities afforded by the mode design in active magneto-optics for controlling the light polarization. Finally, in the T-MOKE (Figure 4b, bottom) two directly excited plasmon nanoantenna modes $\alpha_{xx}$ and $\alpha_{zz}$ are spin-orbit coupled. The Kerr polarization rotation is detected as the T-MOKE contrast, which is the ratio between the difference in intensity of the reflected light for both magnetization directions and the reflected light intensity in the non-magnetized state ($\Delta I / I_0$). As the $p_y$ dipole does not contribute, the nanoantenna modes design plays even larger role here. The black and red configurations show relatively low contrast for long wavelengths ($\alpha_{xx}$ off-resonant) compared to the green and blue ones ($\alpha_{xx}$ and $\alpha_{zz}$ resonant). At shorter wavelengths, we see the radically intensified contrast when approaching the $\alpha_{xx}$ resonance (red configuration, for example). Although in T-MOKE we do not alter the polarization state as such, it proves to be a practical experimental geometry for the amplitude modulation of the reflected light. 

We set the generalized analytical expressions for the Kerr polarization rotation of Figure 4b  to be applied as design rules for magnetoplasmonic systems in which unprecedented control of the polarization state of light is achieved. By engineering the nanoantenna elements along all three symmetry axes the polarizabilities can be tailored at will, resulting in broadband spectral control. Furthermore, the incident angle brings an additional degree of freedom as it can be used to control the coupling efficiency to the two direct nanoantenna dipoles, and thus also to the spin-orbit induced dipoles. 

In the scenarios above we utilize p-polarized light, as we can directly couple to two of the nanoantenna modes (with off-normal incidence), and we invoke the third nanoantenna mode via spin-orbit coupling. For s-polarized light (see Supporting information), we only obtain direct coupling to one nanoantenna mode ($\alpha_{yy}$) for the three MOKE geometries. This implies that in s-polarized L-MOKE and P-MOKE we couple through the spin-orbit interaction to the nanoantenna modes $\alpha_{zz}$ and $\alpha_{xx}$ respectively, while in T-MOKE the only involved mode is the one that is directly excited. Essentially, for s-polarized light the L- and P-MOKE polarization rotations are defined by one spin-orbit induced nanoantenna dipole mode, resulting in a rather limited polarization tunability (see Supporting information for the example of the 2D nanoantenna in L-MOKE).

Summarizing, we experimentally and theoretically verify the emergence of active magneto-optics, where magnetoplasmonic nickel nanoantennas control the polarization states (rotation and ellipticity) of the reflected light in a broadband regime. The general concept applies to magnetoplasmonic nanoantennas with various mode design and can be implemented for all conventional Kerr polarization rotation geometries, namely L-MOKE, P-MOKE and T-MOKE. We derived analytical expressions with a predictive power for the polarization rotation in all mentioned experimental geometries. This generality prompts the use of the developed design principles for the emerging magnetoplasmonic metasurfaces that would deliver unprecedented control over the polarization state of light by means of magnetic fields in a whole range of future active nano-optic devices. 

{\bf Supporting information: } Spectroscopic L-MOKE spectra for 2D nanoellipses in s-polarization. Overview of the design rules and their derivation for the three conventional MOKE geometries in p- and s-polarization. Calculated MOKE response for the various geometries in 2D and 3D nanoellipses, including different incident angles. This material is available free of charge via the Internet at http://pubs.acs.org.

{\bf Acknowledgements: } K. L., I. Z. and A. D. acknowledge funding from the Swedish Research Council (VR), Swedish Foundation for Strategic Research (SSF) and Chalmers Area of Advance Materials Science. N. M. and P. V. acknowledge funding from the Basque Government (Program No. PI2012-47 and Grant No. PRE-2013-1-975) and the Spanish Ministry of Economy and Competitiveness (Project No. MAT2012-36844). R. K. D. and J. \AA. acknowledge funding from the Swedish Research Council (VR), Swedish Foundation for Strategic Research (SSF) and the Knut and Alice Wallenberg Foundation.

\end{document}